\documentclass[twocolumn,showpacs,reprint,amsmath,amssymb,eqsecnum,pre,showkeys,aps,superscriptaddress]{revtex4-1}
\usepackage{graphics}
\usepackage[caption=false]{subfig}
\usepackage{graphicx}
\usepackage{bm}
\usepackage[usenames]{color}
\newcommand{\beq}{\begin{equation}}
\newcommand{\eeq}{\end{equation}}
\newcommand\beqa{\begin{eqnarray}}
\newcommand\eeqa{\end{eqnarray}}

\newcommand{\ed}{\bibliographystyle{apsrev}\bibliography{myBIB}\end{document}}

\begin{document}
\title{Distributed Computing on Complex Networks}
\author{Francisco Prieto-Castrillo}
\email{francisco.prieto@ciemat.es}
\affiliation{Research Centre for Environment, Energy and Technology, Spain}
\author{Antonio Astillero}
\email{aavivas@unex.es}
\affiliation{Department of Computer Science and Technology, University of Extremadura, Spain}
\author{Mar\'{i}a Bot\'{o}n-Fern\'{a}ndez}
\email{maria.boton@ciemat.es}
\affiliation{Research Centre for Environment, Energy and Technology, Spain}
\date{\today}
\begin{abstract} 
This work considers the problem of finding analytical expressions for the expected values of distributed computing performance metrics when the underlying communication network has a complex structure.
Through active probing tests a real distributed computing environment is analysed. From the resulting network, ensembles of synthetic graphs with additional structure are used in Monte Carlo simulations to both validate analytical expressions and explore the performance metrics under different conditions. Computing paradigms with different hierarchical structures in computing services are gauged, fully decentralised (i.e., \textit{peer-to-peer}) environments providing the best performance.
Moreover, it is found that by implementing more \textit{intelligent} computing services configurations (e.g., betweenness centrality based mappings) and task allocations strategies, significant improvements in the parallel efficiency can be achieved. We qualitatively reproduce results from previous works and provide closed-form solutions for the expected performance metrics linking topological, application structure and allocation parameters when job dependencies and a complex network structure are considered.
\end{abstract}
\pacs{89.75.-k,  05.10.Ln, 89.20.Ff, 02.50.-r}
\keywords{Complex networks, distributed computing, Grid}
\maketitle
\section{Introduction\label{sec:INTRO}}
As emphasised by L\'{a}szl\'{o} Barab\'{a}si~\cite{Barabasi:2009vi}, once the universal properties of complex networks have been identified, the challenge is to correlate those properties with the processes taking place on the networks. In this regard, an unsolved problem is the effect of the topological characteristics of the underlying communication network on the performance of a distributed computing infrastructure.\par
From a top-level view of computer function and interconnection, Distributed Computing Infrastructures (DCI) can be understood as the coordination of computing and storage resources through a communication network aimed at solving a task. Depending on the network span (local or global), business model for user-system interactions and the hierarchical structure of deployed services, different DCI solutions have emerged in the last decades: \textit{Grid, Cloud, P2P, Cluster, Utility, Volunteer, Parasitic} and \textit{Jungle computing}. In Ref.~\cite{Kahanwal:uq} a review and classification of these paradigms can be found.\par
Regarding the network scale, DCIs can be split into two main categories: \textit{Cluster Computing}, where nodes are usually connected through high bandwidth and low latency (usually fibre channel) links,  and the remaining solutions listed above, which involve a geographically dispersed network (e.g., the internet).\par 
From the business model point of view three paradigms coexist; \textit{utility computing, volunteer computing} and \textit{parasitic computing}. In the \textit{utility} model users pay for on-demand delivery of computational or storage services and the infrastructure operates in a transparent manner to the user by aggregating resources in a single system view. This includes Grid and Cloud computing. Grids can be understood as federations of computing clusters that coordinate their resources for a specific scientific community (\textit{Virtual Organisation} or \textit{VO})~\cite{Foster:2001th}. Within this federation each cluster is autonomous in their access policies and in the amount and type of resources they offer. Furthermore, resources can be switched on and off dynamically without a centralised control. Grids are often deployed on LAN, WAN, or internet backbone networks at regional, national, or global scales. Cloud is the latest evolution from Grid (still under development) which by means of an intensive use of virtualization technology can offer different computing capacities as a service, usually in a pay-per-use model. There are many similarities among Grid and \textit{Cloud Computing} (for a complete survey see Ref.~\cite{Stanoevska-Slabeva:1315759}). In a rough way it can be stated that Cloud relies on \textit{Grid Computing} as its backbone and infrastructure support.
\par
\textit{Volunteer computing} is another collaboration scheme in which users altruistically donate their computing capacities to a project. A problem is split into tasks that can be evaluated independently by computers connected to the internet. Those tasks (\textit{work units}) are then collected by a central server. A celebrated example is the SETI@home project (see http://www.seti.org). At the opposite end of the spectrum lies the \textit{parasitic computing} solution~\cite{Barabasi:2001vt}, which retrieves computing resources without the knowledge of the participating servers.
\par
Large-scale, geographically dispersed DCIs can be classified according to the hierarchical relationships among their components. In the Grid case, the task mapping to clusters (\textit{Computing Elements} or simply \textit{CEs}) is coordinated by \textit{meta-schedulers} (\textit{Resource Brokers} or simply \textit{brokers}) acting as servers. CEs are responsible for managing the nodes (\textit{worker nodes} in the Grid terminology) where tasks are finally executed. \textit{Brokers} find CEs according to user specifications and CE's performance evaluation, availability and other rankings~\cite{Foster:2001th}. In this regard, if two tasks hosted in two CEs need to communicate, a path passing through the involved \textit{brokers} is created. Depending on the number of \textit{brokers} the system can be more or less hierarchical; from a totally centralised solution (single \textit{broker}) to a decentralised solution in which there is a \textit{broker} for each CE connected to the same communication node in the network. Cloud virtualization layers  introduce additional complexity and the component schemes can be very intricate as the \textit{elastic} nature of \textit{clouds} enables the change of resource quantities and characteristics at runtime. However, the \textit{Cloud WorkFlow Management System} (WFMS) still makes use of schedulers and worker nodes in a hierarchical way~\cite{Buyya:2011vi}. \textit{Peer-to-peer computing} (or simply P2P) lies at the opposite end. The P2P solution replaces the distinct notions of server and client nodes with the notion of peers~\cite{Karagiannis03filesharingin}. Hence, P2P systems result in a fully distributed configuration where nodes (i.e., \textit{peers}) are connected without any other intermediary service. P2P can be thought as the limit case of total decentralisation in Grids. Notice also that volunteer computing can also be thought as the total centralised case of Grids.
\par
Finally, these computing paradigms can be more or less homogeneous in their resources. Recently, the trend appears to be the merging of several solutions resulting in a highly heterogeneous scheme known as \textit{jungle computing}~\cite{Kahanwal:uq}.\par
The understanding and design of DCIs with nontrivial communication topologies and dynamically changing conditions (network traffic, node connections, etc.) demands new methodologies and tools different from those usually applied in High Performance Computing~\cite{Muttoni:2004us}. In fact, modelling the interactions among users, application structures (workflows) and DCIs continues to pose a big and ongoing challenge.  A promising approach is the integration of concepts and tools from the complex systems theory into DCI models. However, few contributions have tackled the task allocation problem in DCIs from that perspective~\cite{Iamnitchi:2002wx,daFontouraCosta:2005wn,Ilijasic:2009wp,Ishii_acomplex}.
\par
In 2002 Iamnitchi \textit{et al.}~\cite{Iamnitchi:2002wx} showed how P2P scientific collaboration networks exhibit the small-world property.  This work represents one of the first contributions of complex network theory to the DCI task allocation problem through the user-infrastructure interaction standpoint. The highly inspiring article by da Fontoura \textit{et al.} in 2005~\cite{daFontouraCosta:2005wn} analysed the effect of a complex network topology in the efficiency of computing Grids for the first time. There, a homogeneous network with constant latency and no background traffic was considered. In their model the authors used undirected graphs with random, scale-free and several customised topologies with nodes representing processing units and edges representing communication links. In this scheme every node was able to compute tasks and to forward tasks to its neighbouring nodes within its cluster until the whole task set was completed. However, the applications considered were restricted to sets of independent jobs with no dependence relationships (i.e., lacking any structure). Analytical relationships between the main performance metrics and topological parameters were not addressed.
\par
An alternate workaround is to tackle the DCI scheduling through the queueing networks formalism (Muttoni \textit{et al.}~
\cite{Muttoni:2004us}) or by an entropy-based scheduling approach (Derbal~\cite{Derbal:2006wu}). In the latter case the capacity of a given service was modelled as a Markov chain and the uncertainty on the service capacity information was quantified in terms of an entropy function. In that work the author considered a graph (termed as \textit{Grid Neighbourhood}) as a dynamic federation of resource clusters composed of two node types; \textit{Principals} and \textit{Agents}. \textit{Principals} (associated, at a minimum, with schedulers) managed several Agents (nodes where jobs are processed) forming a cluster and the edges linking the \textit{Principals} defined the Grid topology. Remarkable performance improvements with respect to the random solution were achieved as the number of clusters increased. Again (except for network size), no explicit relationships were reported between performance metrics and topology (claimed as holding the power-law property but without providing any evidence of this finding).
\par
Ishi and colleagues~\cite{Ishii_acomplex} went a step further in 2007 by tackling the scheduling problem in Grids through an optimisation approach using a simulated annealing algorithm. In that model every graph node represented a CE and communication paths were built through the Dijkstra algorithm for a homogeneous network with random, small-world and scale-free structure. As in~\cite{daFontouraCosta:2005wn} the applications lack any dependence structure or analytical relationships between metrics and topological parameters. Also in 2007, Batista \textit{et al.}~\cite{Batista:2007tv} proposed a procedure for enabling Grid networks to dynamically self-adjust to resource availability. Mechanisms for task scheduling, resource monitoring and task migration adaptation were provided. However no reference to either model expressions, parameters or the underlying mechanisms enabling the alleged self-adaptability is found in that work.
\par
More recently Llija\^si\'c and Saitta~\cite{Ilijasic:2009wp} considered user-infrastructure interactions as in~\cite{Iamnitchi:2002wx}; this time through a probabilistic graph model approach. In that work Grid log data of more than 28 million jobs from the EGEE (Enabling Grids for E-sciencE) EU funded project~\cite{Jones:2005fk} were collected and mined. User-Grid processes were modelled through the generation of a directed bipartite graph with nodes representing either users or CEs and links representing users sending tasks to CEs. The resulting graph, with a shortest path length of $3.56$ and a diameter of $10$, rendered a power-law structure in its out-degree distribution.
\par
All these works faced the DCI performance problem through techniques akin to complex systems based methods, either from the infrastructure or from the user-application perspective. However, none of them provides analytical relationships among infrastructure, application and allocation parameters. Although complex network based models were used, no evidence of why these models should reproduce the referred DCI's topology is found. Furthermore, assembling jobs into clusters targeted to the same computing resource (i.e., \textit{job clustering}) has a remarkable effect on the performance, as in this case communication overheads are zeroed. Moreover, well-known workflow management systems and algorithms have been designed and successfully applied to both multiprocessors and distributed computing environments~\cite{McCreary:1994ur,Forti:2006un,Derbal:2006wu,Deelman:2003wk,Cao:2009bw}. However, to the authors' knowledge, no previous works have included intertask dependencies effects when analysing DCIs with a complex network topology.
\par
This paper is aimed at finding functional relationships between application performance metrics and the key parameters of different DCI solutions. Inspired by the topology and job services organisation of a real Grid infrastructure (Sec.~\ref{sec:EXP}), we develop a probabilistic model for application performance metrics in Sec.~\ref{sec:MODEL}. In particular we address models for: (1) Simple application workflow schemes (but with job dependence relationships), (2) Parametrized hierarchical structure among DCI services, (3) Probabilistic job allocations based on task clustering. As a result, expressions of the first moments from the resulting order statistics are obtained. These expressions are then validated through Monte Carlo simulations for different DC solutions and configurations (Sec.~\ref{sec:RESULTS}). We conclude and make some remarks on the scope of the addressed model in Sec.~\ref{concl}.
\section{Network Tomography of a real Grid infrastructure \label{sec:EXP}}
Global DCIs (Grid, Cloud, P2P and \textit{Jungle}) overlay computing, storage and other software services over the internet. In this scheme the resulting network consists on nodes belonging to both services and communication layers in a multiplexed-network fashion~\cite{Foster:2001th}. In this work we start from the study of a real computing Grid to scan its structural properties. In particular we study the Workload Management System (WMS) of the generic VO \textit{Ibergrid} in the joint \textit{Spanish and Portuguese National Grid Initiatives} Grid (ES-NGI) within the European Grid Infrastructure project~\cite{EGI}. This DCI can be understood as a logical network with three node types: \textit{brokers}, computing elements and communication nodes (routers). The resulting topology is determined by the coupling between the internet topology and the \textit{brokers} and CEs mapping mechanisms. Hence, it can be expected that the resulting graph preserves some structural properties from the internet. 
\par
The internet can be viewed as an evolving network of subnetworks (\textit{Autonomous Systems} (AS) or \textit{domains}). Every domain (managed by an administration authority with specific policies) is a network composed by routers, switches and hosts (end computers where application programs run). Hence, the internet topology is modelled at two granularity levels; inter-domain (AS or domain level) and intra-domain (router-level). In Ref.~\cite{Faloutsos:1999tk} it was claimed that power-law relationships exist at both inter and intra domain levels from the BGP routing tables collected by the route server \textit{route-views.oregon-ix.net}. However, this assertion has been a subject of great controversy. For instance, in Ref.~\cite{Chen02theorigin} it is highlighted that derived AS-level topology is not representative for the internet connectivity since at least 20-50\% of the physical links were missing in that work. Further studies suggest that this power-law at the AS-level remains when the AS maps are extended~\cite{Siganos:2003hu}.
\par
The alleged intra-domain power-law relationship in Ref.~\cite{Faloutsos:1999tk} derives from the \textbf{traceroute} (see http://www.caida.org/tools) tests collected in~\cite{Pansiot:1998vo}. This tool allows tracking packet destinations along a path at the IP layer of the internet. By merging all these paths a reconstruction (i.e., \textit{active network tomography}) of the network topology is achieved.  However, this technique revealed several shortcomings for rendering a reliable topology. A common pitfall is to infer the existence of a physical node that is really a logical entity. In Refs.~\cite{Amini2004557,Willinger:2009vj} a survey of biases taking place in \textbf{traceroute} like probes can be found.
\par
In the present work \textbf{traceroute} active probing, similar to that in Ref.~\cite{Pansiot:1998vo}, is used. The graph reconstruction process is as follows: 1) Retrieve IP address of every \textit{broker} and CE, 2) For every CE send an \textit{agent} job with the remaining CEs IP addresses and a \textbf{traceroute} command for those destinations, 3) The resulting set of paths is processed so that each intermediate IP is regarded as a node (it can be a router, a virtual node or even a whole domain) and a link is created for two consecutive IPs. This way local views are obtained by evaluating a certain number of paths from many sources to different destinations. The merging of these views provides a \textit{snapshot} (the Grid is an evolving entity) of a partial map of the DCI underlying communication network. The obtained graph is of size $189$ ($161$ communication nodes, $23$ CEs and $5$ \textit{brokers}) with $477$ edges. An average degree $k=5.03$ and average clustering coefficient $CC=0.106$ are found. The graph diameter and average shortest path computations renders $l=4.24$ and $D=15$ respectively.
\par
With the procedures described in Ref.~\cite{Clauset:2007vw} for dealing with small samples, a power-law fit for an exponent $\alpha=2.78\pm0.13$ for $k_{\text{min}}=5\pm 1$ is obtained. The goodness-of-fit is evaluated through the \textit{plpva} function provided in Ref.~\cite{plfitweb} resulting in a $p-value = 0.1350$. This result, despite the possible \textit{poor statistics} effect due to the small size of the graph, supports the hypothesis that the graph might be scale-free. However from the discussion above we do not claim that the ES-NGI communication network has in fact a power-law structure. In particular, as we could only resolve a subset of domains (i.e., the \textit{IP alias resolution problem}~\cite{Willinger:2009vj}), the resulting graph is undoubtedly incomplete, rendering only a logical representation of the network. This description falls somewhere in between the inter-domain and intra-domain levels~\cite{Siganos:2003hu}. Although being incomplete, this graph still serves as a seed and reference model for the simulations conducted in this work.
\section{A probabilistic model for task allocation in distributed computing environments\label{sec:MODEL}}
\begin{figure}[h]
   		\centering
  		 \includegraphics[width=.9\columnwidth]{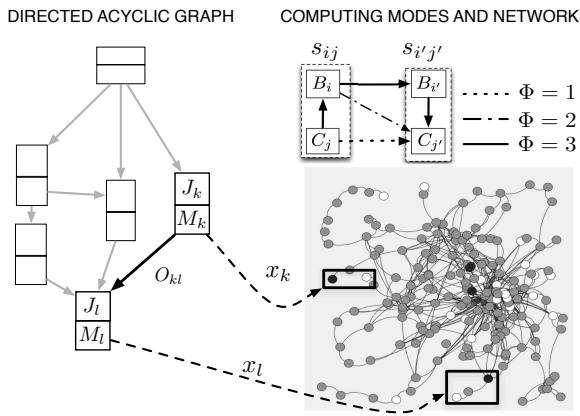}
  		 \caption{Stochastic allocation in a distributed computing infrastructure. The overlay of computing services (\textit{brokers} and CEs represented as black and white circles) deployed on the communication network (routers represented as grey circles) causes different degrees of hierarchy. The overhead transfer costs in every DCI are parametrized through the quantity $\Phi=1,2,3$. The network shown represents a real snapshot built from \textbf{traceroute} experiments on the ES-NGI DCI.}
  		 \label{fig:modes}
	\end{figure}

The stochastic task allocation process can be thought as the coupling of three structures: application workflow, computing network and allocation process (see Fig.~\ref{fig:modes}). In this section these structures are formalised. Then, the first moments of DCI performance metrics are derived from the resulting order statistics. Finally we provide analytic expressions for the limit cases considered.  
	\subsection{Scientific application workflow modelling}
 Scientific applications are typically modelled as workflows, consisting of tasks, data elements, control sequences and data dependencies~\cite{book_task_scheduling}. They can be formalised as weighted Directed Acyclic Graphs (DAGs), $G_{J}=(J,E_{J})$ where $J$ is a set of $n_{J}+1$ Jobs $J=\{J_{0},J_{1},\ldots,J_{n_{J}}\}$ and $E_{J}$ is a set of $ne_{J}$ directed edges $E_{J}\subset J\times J$.
Every job $J_{k}$ has an associated weight $M_{k}$ representing its size (in bytes) and every link $e_{kl}=(J_{k},J_{l})$ carries also a weight $O_{kl}$ representing the file size (in bytes) transferred between jobs $J_{k}$ and $J_{l}$ when a task communication process holds. In order to simplify computations, the special entry-job $J_{0}$ with $M_{0}=0$ and $O_{0k}=0,\forall k$ is introduced without loss of generality~\cite{Forti:2006un}.
\par
\begin{figure}[h]
 \includegraphics[width=.9\columnwidth]{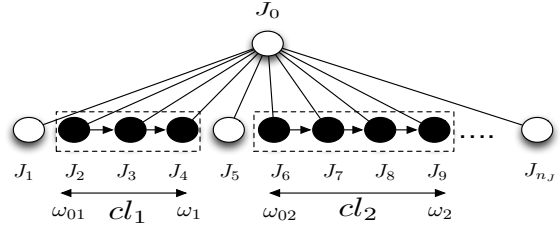}
\caption{Simplified parametrizable DAG model for application workflows used in this work. Dependent jobs are arranged into clusters $cl_{i}$ of size $\omega_{i}-\omega_{0i}$.}
\label{fig:DAG_PARAM}
\end{figure}

DAGs can be parametrized according to different metrics depending on their adjacency matrix $A_{ij}$ and node/link weights. Roughly, the basic DAG components are the fork and join structures. Given a job $J_{i}$, in a fork or join structure the out-degree $k^{out}_{i}$ or the in-degree $k^{in}_{i}$ are non-zero respectively. In this work we assume a minimalistic parametrizable DAG; DAGs are such that $k^{in,out}_{i}\in\{0,1\},\;\forall i\in I_{J}$ and with an entry-node $k^{in}_{0}=0$, $k^{out}_{0}=n_{J}$. This results in a set of linked \textit{job clusters} $cl_{k}$ of different sizes $\omega_{k}-\omega_{0k}$ (see Fig.~\ref{fig:DAG_PARAM}).
Further, we will assume equally sized jobs (i.e., \textit{probe jobs}) $M_{k}=M,\forall k$ and transfer files $O_{kl}=o, \forall k,l$. Although this is a non-negligible constraint for representing real DCIs, using probe jobs allows to explicitly highlight the effects of the network topology on the applications performance, which is the major aim of the present contribution.
	\subsection{Distributed computing infrastructure model}
From the previous discussion in Sec.~\ref{sec:EXP} a node can represent a router, switch or even an AS. Here we overlook that distinction for the purposes of the present model. The adopted graph $G_{R}=(R,E_{R})$ consists on a set $R$ of $n_{R}$ communication nodes (routers) connected by $ne_{R}$ links $r_{ij}\in E_{R}$ characterised by a latency $l_{ij}$ (in seconds) and bandwidth $bw_{ij}$ (in bytes per second), as it is usual in computer networks analysis~\cite{Peterson:ZPL0tymJ}. Moreover, we assume bidirectionally in the network $l_{ij}=l_{ji}$,  $bw_{ij}=bw_{ji}, \forall i,j$ and that it is always possible to find a geodesic path $P_{ij}=(R_{i},\ldots,R_{j})$ of length $d_{ij}$ between any two nodes $R_{i}$ and $R_{j}$. For every path we define both the path latency $L_{ij}=\sum_{(R_{a},R_{b})\in P_{ij}}l_{ab}$ and path bandwidth  $BW_{ij}=\min\{bw_{ab}:(R_{a},R_{b})\in P_{ij}\}$. By considering homogeneous latencies $l_{ij}=LAT$ and bandwidths $bw_{ij}=BW$ in the network, the path quantities render: $L_{ij}=LAT\cdot d_{ij}$ and $BW_{ij}=BW$ respectively. For every graph instance $g\in G_{R}$ we use the following definition for the average shortest path
\begin{equation}
l(g)=\frac{1}{\binom{n_{R}}{2}}\sum_{i\neq j}d_{ij}(g),
\label{eq:l}
\end{equation}
and $D(g)=max\{d_{ij}(g)\}$ for the network diameter $D(g)$. From the inter-router distance distribution $f_{d}(x)=P[d\leq x]$ (with the random variable $d$ as the geodesic distance between two randomly chosen nodes the graph) and the \textit{tail} distribution $\tilde{f_{d}}=P[d>x]$, equation~(\ref{eq:l}) can also be expressed as
\begin{equation}
l(g)=\sum_{x=0}^{D(g)-1}\tilde{f}_{d}(x).
\label{eq:l_tail}
\end{equation}
Regarding DCI services, a simple model for workload allocation is assumed. In particular, we do not consider other services such as storage, information, resources discovery, etc. In a task communication process between jobs $J_{k}$ and $J_{l}$ with a dependence relationship ($A_{kl}\neq 0$) the communication overhead depends on the DCI solution adopted. In the P2P case CEs transfer files directly (see Fig.~\ref{fig:modes}). However, one or more meta-schedulers (\textit{brokers}) can also be involved in the process. In the Grid case it is usual to find a hierarchical structure where a single \textit{broker} covers a wide range of CEs. Other \textit{brokers} can also be used as a support for redundancy, communication optimisation, failure, or by direct user specifications.
\par
We then characterise \textit{distributed computing configurations} through the constant $\Phi={1,2,3}$. These solutions depend on the number of \textit{brokers} and CEs deployed (denoted as $n_{B}$ and $n_{C}$ respectively) and their connections (Fig.~\ref{fig:modes}). With a slight abuse of the $\delta$ notation we define $\Phi=3-\delta_{1n_{B}}-2\delta_{P2P}$ where $\delta_{P2P}=1$ if $n_{B}=n_{C}$ and both \textit{brokers} and CEs share the same router (otherwise  $\delta_{P2P}=0$). This parametrizes DCI solutions as follows: 1) P2P with $\Phi=1$ and $n_{B}>1$ (fully decentralised solutions). 2) Total centralisation with $\Phi=2$, where $n_{B}=1$ and at least one CE is connected to a different router. 3) Partially centralised solutions with $\Phi=3$ for $n_{B}>1$ and where at least one CE is connected to a different router than those connected to the \textit{brokers}.
\par
For every DCI we also define: 1) A set $S=B\times C$ of $n_{S}$ computing modes composed by a \textit{broker} set $B=\{B_{1},\ldots,B_{n_{B}}\}$ and a CE set $C=\{C_{1},\ldots,C_{n_{C}}\}$, 2) Two simple  (with finite support) random variables $\beta: B\rightarrow R, B_{i}\mapsto \beta_{i}$ and $\theta: C\rightarrow R,C_{j}\mapsto \theta_{j}$ to model broker and CEs services mappings respectively. Computing modes are denoted as $s_{ij}=(B_{i},C_{j})$ but in order to lighten notation we eventually set a single index labelling through the bijection $(i,j)\mapsto n_{C}(i-1)+j; i=1,\ldots,n_{B},j=1,\ldots,n_{C}$. 
\par
In our model we consider CEs with infinite processing capacity, no queuing delays and equal processing speeds $V$  -measured in Million of Instructions Per Second (MIPS)-. Infinite processing capacity means that if two tasks arrive at the same CE they are executed in parallel if no precedence relationship exists in the corresponding DAG. This takes our approach slightly away from the pure scheduling models where every task has an assigned order; if two tasks with no precedence relationship arrive at the same host the local scheduler establishes an additional order of execution~\cite{book_task_scheduling,Derbal:2006wu,Forti:2006un,Batista:2007tv,Zhang:2011vl,Ishii_acomplex}. In our approach it suffices to comply with the order imposed by the DAG. The effect is equivalent to CEs with infinite processing capacity so that no additional ordering is required. Furthermore, \textit{brokers} are assumed to hold infinite storage capacity and task forwarding speed; there is no congestion and tasks are efficiently dispatched without introducing any additional delays.
	\subsection{Allocation process and time metrics}
Task allocation is modelled by the stochastic process $AP=\{X_{k}:k\in I_{J}\}$, where $I_{J}$ denotes the job index set and every random variable $X_{k}$ has support $S$ and $x_{k}\in S$ represents the assigned mode to job $J_{k}$ (Fig.~\ref{fig:modes}). We also define the function: $\Delta:S\times S\rightarrow\{1,0\},(x_{k},x_{l})\mapsto \Delta_{kl}$, where $\Delta_{kl}=1$  if $x_{k}=x_{l}$ and $\Delta_{kl}=0$ if $x_{k}\neq x_{l}$. Mass probabilities $P[X_{k}=s_{n}]$ are denoted as $p_{kn}$, joints  $P[X_{k}=s_{n},X_{l}=s_{m}]$ as $p_{kn,lm}$ and conditionals $P[X_{k}=s_{n}|X_{l}=s_{m}]$ as $p_{kn|lm}$. With these elements, a probabilistic model for the conditional probabilities is built (see details in Appendix~\ref{ap0}). The aim is to provide independence between the events $"\{x_{k}=s_{n}\}"$ and $"\{x_{l}=s_{m}\}"$ when there is no order relationship between jobs $J_{k}$ and $J_{l}$ ($A_{kl}=0$) while parametrizing the strength of the likelihood of sharing a mode when an order dependency holds. Through a simple model~(\ref{eq:P_delta}), the likelihood of two jobs $J_{k}$ and $J_{l}$ for arriving at the same mode $\tilde{p}_{kl}$ can be parametrized as follows: when there is an order relationship in the DAG, $\tilde{p}_{kl}=c$, where $c\in[0,1]$ is the \textit {task clustering parameter}, and when that relationship is missing the mapping is simply uniformly random $\tilde{p}_{kl}=1/n_{S}$. This way, $c$ controls the likelihood of two adjacent jobs for being mapped to the same mode (i.e., zeroing communication costs). This process is termed as \textit{job clustering}~\cite{Forti:2006un}.
\par
As no queueing delays and homogeneous processing speeds are considered, processing times are also constant and independent of the mapping. Hence, job processing times $PT$ (in seconds) are simply defined  as $PT_{k}=M/V,\;\forall k\in I_{J}$. Further, we will only consider the case of  \textit{latency-bounded applications}, where  the ratio $o/BW<<LAT$. This can be achieved either with small file sizes or with a high bandwidth network (as it is usual in dedicated research networks and many computing Grids). In the experimental study of the ES-NGI Grid described in Section~\ref{sec:EXP} Round Trip Times (RTT) for every IP packet were also collected. We found a minimum $BW=$48.32Kbytes/s and a maximum latency $\sim10^{-1}$s. Hence, the latency-bounded approximation is justified for file sizes of the order of bytes.
\par
Another simplification is achieved if equally sized clusters are considered. Then, communication costs between jobs $J_{k}$ and $J_{l}$  can be computed as
\begin{equation}
CT_{kl}=[1-\Delta_{kl}][LAT\cdot D_{kl}^{cc}],
\label{eq:CT}
\end{equation}
where $D_{kl}^{cc}=D_{k}^{cb}+D_{kl}^{bb}+D_{l}^{bc}$ is the inter-CE joint distance (see Fig.~\ref{fig:modes}) composed of: 1)  Intra-mode \textit{CE-broker} distance $D_{k}^{cb}$ at the departure mode $x_{k}$. 2) Inter-mode \textit{broker-broker} distance $D_{kl}^{bb}$ from modes $x_{k}$ to $x_{l}$. 3) Intra-mode \textit{broker-CE} distance $D_{l}^{bc}$  at the arriving mode $x_{l}$. It must be highlighted that this linear relationship (also used in~\cite{daFontouraCosta:2005wn}) between communication times and hosts distance is only a rough approximation to the real internet communication processes; as communications depend heavily on network congestion (i.e., traffic) the assumption of a constant latency may not be adequate in general. On the other hand, a strong correlation has been found between hop-count and packet transfer times~\cite{Fujii:2000td} suggesting that packet transfer times may increase linearly with the distance. This evidence empirically supports the linear relation~(\ref{eq:CT}) that can be thought as a limit case when $O_{kl}/BW\rightarrow 0$ and no background traffic effect is included.
\par
Under these assumptions we now define the set of metrics for the quantification of applications performance used in this work. From the adjacency matrix $A$, the finish time $FT_{k}$ of job $J_{k}$ is given by
\begin{equation}
FT_{k}=PT+\max\{A_{ik}(FT_{i}+CT_{ik})\}_{i=0,\dots,n_{J}}.
\label{eq:FT}
\end{equation}
As for the entry node it holds that $FT_{0} = 0$, and by noticing that for the proposed DAG $A$ is upper-triangular, a perfect recurrence relation is found: $FT_{k}=PT+\max\{A_{ik}(FT_{i}+CT_{ik})\}_{i<k},\;k=1,\ldots,n_{J}$.
\par
The first performance metric considered is the total execution time of a parallel application \textit{makespan} (or \textit{MK} for short) defined as $MK = \max\{FT_{k}\}_{k=1,\ldots,n_{J}}$ (in seconds). 
\par
Another common measure is the \textit{Scheduled Length Ratio} $SLR=MK/CPIC$, which normalises \textit{makespan} with the \textit{Critical Path Length Including Communication} ($CPIC$)~\cite{Forti:2006un,Zhang:2011vl}. $CPIC$ is computed in the same way as $MK$ but in this case tasks are mapped to an \textit{average mode} $\bar{S}$ where both intra-mode and inter-mode distances are equal to the average shortest path length $l$. According to this, the $CPIC$ processing and communication times are defined as: $PT_{k}^{CPIC}=M/V=PT$ and $CT_{kl}^{CPIC}=[1-\Delta_{kl}][\Phi\cdot l\cdot LAT]$ respectively. Then, $CPIC$ is obtained as $CPIC = \max\{FT_{k}\}_{k=1,\ldots,n_{J}}$ by plugging $PT_{k}^{CPIC}$ and $CT_{kl}^{CPIC}$ into~(\ref{eq:FT}). A value of $SLR=1$ means that the allocation achieves a performance equal to that obtained in the \textit{average network}. Values lower than $1$ indicate a better mapping whereas values higher than $1$ render a worse strategy. In general, the lower $SLR$ the better performance obtained. In a uniform mapping $SLR$ has a lower-bound of $1$ but, as it will be shown, \textit{smarter} mappings (i.e., taking into account the services degree or betweenness) can achieve values with $SLR<1$.
\par
Other common metric is the \textit{parallel efficiency} PE or \textit{normalised speedup}. As defined in Ref.~\cite{daFontouraCosta:2005wn} PE is the normalised ratio between sequential and parallel time for an application to execute $PE=\sum_{i}PT_{i}/(n_{S}MK)=n_{J}PT/(n_{S}MK)$.
	\subsection{Probabilistic modelling and order statistics}
The proposed model has three main elements; network, computing services mappings and job allocation process $AP$.
Starting from the sample space $\Omega=G_{R}\times R^{n_{B}}\times R^{n_{C}}\times S^{n_{J}}$, each possible configurations $\textbf{x}\in\Omega$ with  $\textbf{x}=(g,\bar{\beta},\bar{\theta},\bar{X})$ represents; a graph instance $g$, a \textit{brokers} $\bar{\beta}=(\beta_{1},\beta_{2},\ldots,\beta_{n_{B}})$ and CEs mapping $\bar{\theta}=(\theta_{1},\theta_{2},\ldots,\theta_{n_{C}})$ and an allocation vector: $\bar{X}=(X_{1},X_{2},\ldots,X_{n_{J}})$.  This way, any time function $\hat{f}$ is regarded as a simple random variable with expectation $E[\hat{f}]=\sum_{\textbf{x}\in \Omega}\hat{f}(\textbf{x})P(\textbf{x})$. Graph instances are referred to any of the two classical random graph models $G_{R}\in\{G_{n_{R},m},G_{n_{R},p}\}$ as defined in~\cite{Bollobas:1998vw} while $E[]$ and $P()$ denote expectation and probability in the $G_{n_{R},m}$ and $G_{n_{R},p}$ models interchangeably.
\par
Through a simple re-ordering of job indexes and by assuming equally sized job clusters $\omega_{k}-\omega_{0k}=\omega,\forall k$ in the DAG shown in Fig.~\ref{fig:DAG_PARAM}, it is found that
 \begin{equation}
 MK=\omega PT+\max\{\sum_{i=\omega(k-1)+1}^{k\omega-1}CT_{ii+1}\}_{k=1\ldots c_{J}}
 \label{eq:mk_DAG_PARAM}
 \end{equation}
(an equivalent expression is obtained for $CPIC$) that can be rearranged as
\begin{eqnarray}
MK=\omega PT+LAT\cdot Y\\
SLR=\frac{\omega PT+LAT\cdot Y}{\omega PT+LAT\cdot \Phi\cdot l\cdot Z},
\label{eq:SLR_MODEL}
\end{eqnarray}
where the quantities $Y=max\{U_{k}:k=1\ldots c_{J}\}$ and $Z=max\{W_{k}:k=1\ldots c_{J}\}$ are defined through the relations
\begin{eqnarray}
U_{k}=\sum_{i=\omega(k-1)+1}^{k\omega-1}[1-\Delta_{ii+1}]D^{cc}_{ii+1}\label{eq:U}\\
W_{k}=\sum_{i=\omega(k-1)+1}^{k\omega-1}[1-\Delta_{ii+1}].
\label{eq:W}
\end{eqnarray}
\par
Since the expressions for $Y$ and $Z$ involve the maximum funcion, obtaining their expected values is not straightforward and requires techniques from order statistics analysis. By series expansion up to second order of~(\ref{eq:SLR_MODEL}) around $E[MK]$ and $E[CPIC]$ (see Ref.~\cite{kendall1945advanced}) and by neglecting terms $O^{2}(LAT)$, $E[SLR]$ can be approximated by
\begin{equation}
E[SLR]\approx\frac{\omega PT+LAT\cdot E[Y]}{\omega PT+\Phi\cdot E[l]\cdot LAT\cdot E[Z]},
\label{eq:E_m_SLR}
\end{equation}
where we have used that $l$ and $Z$ are independent variables since $l$ depends solely on topology while $Z$ depends on allocation. The quantities $E[Y]$ and $E[Z]$ can be computed by using the conditional expectation rule (the corresponding partition in $\Omega$ is always possible): $E[Y]=\sum_{g\in G_{R}}P(g)E[Y|g]$ (same for $Z$) where $P(g)=P[\{g\}]$ and $E[Y|g]$ represent the conditional expectation of $Y$ for  a given graph instance $g$. By assuming that both $U_{k}$ and $W_{k}$ with $k=1,\ldots, c_{J}$ form Independent and Identically Distributed (IID) sets of random variables, the set $\{U_{1},\ldots,U_{c_{J}}\}$ (same for $W$) can be thought as a set of statistical samples from the simple random variables $U:\{g\}\times R^{n_{B}}\times R^{n_{C}}\times S^{\omega-1}\rightarrow\mathbb{N}$ and $W:S^{\omega-1}\rightarrow\mathbb{N}$ with distributions $f_{U}(x)$ and $f_{W}(x)$ respectively. Hence, the first moments of  $Y$ and $Z$ can be computed form the $c_{J}$-th order statistics of  $U=\sum_{i=1}^{\omega-1}(1-\Delta_{ii+1})D_{ii+1}^{cc}$ and $W=\sum_{i=1}^{\omega-1}(1-\Delta_{ii+1})$ with supports $\{0,1,\ldots,S_{U}\}$ and $\{0,1,\ldots,S_{W}\}$ respectively. By using the results reported in~\cite{Arnold:2008uc} it is found that
\begin{eqnarray}
E[Y|g]=S_{U}-T_{U}(S_{U},c_{J})
\label{eq:E_p_Y}
\\
E[Z|g]=S_{W}-T_{W}(S_{W},c_{J}),
\label{eq:E_p_Z}
\end{eqnarray}
where the quantities: $T_{U}(S_{U},c_{J})=\sum_{x=0}^{S_{U}-1}[f_{U}(x)]^{c_{J}}$ and $T_{W}(S_{W},c_{J})=\sum_{x=0}^{S_{W}-1}[f_{W}(x)]^{c_{J}}$ have been introduced. By noticing that the value $S_{W}$ is reached when $\Delta_{ii+1}=0, \forall i$, it follows that $S_{W}=\omega-1$. Further, as these quantities do not depend on the network
\begin{equation}
E[Z]=E[Z|g]=(\omega-1)-\sum_{x=0}^{\omega-2}[f_{W}(x)]^{c_{J}}.
\label{eq:E_p_Z_ARNOLD}
\end{equation}
\par
The quantity $S_{U}$ can be computed by realising that the compound distances $D^{cc}_{ii+1}$ in ~(\ref{eq:U}) can be rearranged as $D_{ii+1}^{cc}=(1-\delta_{1n_{B}})D_{ii+1}^{bb}+(1-\delta_{P2P})(D_{i}^{cb}+D_{i+1}^{bc})$.  As the services can be mapped to any node in the network, the support of $D_{i}^{cb}$, $D_{ii+1}^{bb}$ and $D_{i+1}^{bc}$ is $\{0,\ldots,D\},\forall i\in I$. This way, the support of $D_{ii+1}^{cc}$ is $\{0,1,\ldots,\Phi D\},\forall i\in I$. Hence, as task clustering and topology are considered as independent, it is obtained that $S_{U}=\Phi DS_{W}$, which couples DCI solution $\Phi$, network diameter $D$ and task clustering $W$. With these values~(\ref{eq:E_p_Y}) renders
\begin{equation}
E[Y]=\Phi E[D](\omega-1)-\sum_{g\in G_{R}}P(g)T_{U}(\Phi D(\omega-1),c_{J}).
\label{eq:Ecompuesta}
\end{equation}
A series expansion of  $T_{U}$  around  $E[S_{U}]$ in the sum of~(\ref{eq:Ecompuesta}) leads to
\begin{eqnarray}
\nonumber E[T_{U}(S_{U},c_{J})]\approx T_{U}(E[S_{U}])+\\
\nonumber\frac{1}{2}Var(S_{U})\partial^{2}_{z}T_{U}(z,c_{J})|_{z=E[S_{U}]},\\
\label{eq:EDlimit}
\end{eqnarray}
where $Var(S_{U})=\Phi(\omega-1)Var(D)$.
\par
In our analytical model the diameter variance is neglected (reasons for this will be provided later). Moreover, as $D$ can only take positive integer values, considering $E[D]$ introduces round-off biases. A way to tackle this is by setting  $\tilde{D}=round(E[D])$ where the $round$ function outputs the nearest natural number of $E[D]$. Then from~(\ref{eq:Ecompuesta}) and~(\ref{eq:EDlimit})
 \begin{equation}
E[Y]=\Phi\tilde{D}(\omega-1)-\sum_{x=0}^{\Phi\tilde{D}(\omega-1)-1}[f_{U}(x)]^{c_{J}}
\label{eq:E_p_Y_ARNOLD}
\end{equation}
and the expressions for the first moments of $MK$ and $CPIC$ result in
\begin{eqnarray}
\nonumber E[MK]=\omega PT +\\
LAT[\Phi\tilde{D}(\omega-1)-\sum_{x=0}^{\Phi\tilde{D}(\omega-1)-1}(f_{U}(x))^{c_{J}}]
\label{eq:MK_GENERAL}
\end{eqnarray}
and
\begin{equation}
E[CPIC]=\omega PT+\Phi\cdot LAT\cdot E[l][(\omega-1)-\sum_{x=0}^{\omega-2}[f_{W}(x)]^{c_{J}}].
\label{eq:CPIC_GENERAL}
\end{equation}
\par
These relations depend on DAG ($\omega$, $c_{J}$), computing network ($f_{U}(x)$, $l$, $\tilde{D}$, $\Phi$) and allocation ($c$,$f_{W}(x)$) parameters. It can be noticed how $E[MK]$ increases with processing time, cluster size, latency, DCI solution and network diameter while it decreases with clustering probability and with the number of task clusters. The summation term in~(\ref{eq:MK_GENERAL}) and~(\ref{eq:CPIC_GENERAL}) captures the effect of the distance distribution in the underlying topology. The minimum is achieved for configurations such that $f_{U}(x)=1$, (e.g., the complete graph limit) where $E[MK]=\omega PT$. Conversely, if  $f_{U}\approx 0$, $MK$ reaches its maximum value at  $E[MK]=\omega PT+LAT\cdot\Phi\cdot\tilde{D}(\omega-1)$.
	\subsection{Limit cases\label{MODEL_VALID}}	
Despite the simplifications introduced so far, the handling of the expressions for $E[MK]$ and $E[SLR]$ from~(\ref{eq:E_m_SLR}),~(\ref{eq:MK_GENERAL}) and~(\ref{eq:CPIC_GENERAL}) quickly becomes unmanageable. These expressions can be substantially simplified if some additional restrictions are imposed. In particular (see Appendix~\ref{ap} for a more detailed description) we assume
\begin{enumerate}
\item \textit{Job pairs} limit. In this case we only consider task clusters of size $\omega=2$.
\item Random distribution of resources. Both \textit{brokers} and CEs are uniformly mapped over the network with probability $1/n_{R}$. 
For P2P ($\Phi=1$) configurations CEs and \textit{brokers} in a computing mode are mapped to the same router.
\end{enumerate}
Under these assumptions, two limit cases are found: 1) \textit{Total task attraction} $c=1$, where the clustering probability reaches its maximum and all tasks are mapped to the same mode (a rather trivial case) and 2) \textit{Task repulsion} $c=0$;  dependent jobs tend to avoid being mapped to the same mode (a sort of load balancing solution) and tasks are mapped in a \textit{RoundRobin} fashion. In the latter case, if a single cluster is also considered (i.e., $c_{J}=1$) expressions for the metrics are greatly simplified. Finally, if the distance distribution is regarded as uniform and the number of task clusters is large, further simplifications in the metrics expressions are obtained. All these approximations are collected in Table~\ref{tab:limits}. 
	\begin{table}
	\begin{ruledtabular}
	\begin{tabular}{lllr}
								& $c=1$ & \multicolumn{2}{c}{$c=0$}\\
	\cline{3-4}
	Normalised metrics 						& 		& \multicolumn{2}{l}{$c_{J}=1\;\;\;\;\;\;\;\;\;\;\;\;\;\;\;\;$}{$c_{J}>>1$}\\
	\hline
	$(E[MK]-2PT)/LAT$ 	&$0$ & \multicolumn{2}{l}{$\Phi E[l]$}{$\Phi\tilde{D}$}\\
	$E[SLR]$ 	&$1$ & \multicolumn{2}{l}{$1$}{$\frac{1+\Phi\eta\tilde{D}}{1+\Phi\eta E[l]}$}\\
	$E[PE]$ 	&$c_{J}/n_{S}$ & \multicolumn{2}{l}{$\frac{1/n_{S}}{1+\Phi\eta E[l]}$}{$\frac{c_{J}/n_{S}}{1+\Phi\eta\tilde{D}}$}\\
	\vspace{0.01cm}\\
	\hline
	Case	& I & \multicolumn{2}{l}{II}{III\footnote{Considering uniform topologies limit}}\\
	\end{tabular}
	\end{ruledtabular}
	\caption{Performance metrics where simple relations are found for three limit cases of the parameters involved. The dimensionless parameter $\eta=LAT/2PT$ is also introduced.}
	\label{tab:limits}
	\end{table}	
Notice that under these circumstances the normalised performance metrics only depend on three quantities: network distance distribution (topology) $f_{d}(x)$, job clustering probability $c$ and clusters number $c_{J}$.
\section{Model validation and main results\label{sec:RESULTS}}
In this section we numerically investigate the relationships between performance metrics and network topology, DAG structure, and allocation. A set of Monte Carlo simulations consisting on $N$ independent samples for each configuration $\mathbf{x}\in\Omega$ were generated. Then, values for the performance metrics introduced in Sec.~\ref{sec:MODEL} were computed. This was performed in a specifically designed software tool: \textit{Stochastic Grid Workbench} (SGW).
\par
Although a number of popular Grid and Cloud simulators exist (e.g., \textit{GridSim} for Grids~\cite{Buyya:2002vv}), the implementation of customised complex topologies and parametrized DAG structures in these frameworks is not a trivial issue as these environments are mainly designed for general purpose configurations. This motivated the development of SGW, a JAVA based tool which uses the Java Universal Network/Graph Framework API~\cite{citeulike:1567700} to numerically validate the expressions obtained. In SGW graphs can be: 1) Loaded (as the empirical ES-NGI obtained through the procedures described in section~\ref{sec:EXP}) 2) Enriched from a \textit{seed} graph by the random addition of new links,  and  3) Generated through different graph generation algorithms provided in~\cite{citeulike:1567700}. In the case of 1) the graph instance $g\in G_{R}$ in $\mathbf{x}$ is constant and hence $E[l]=l$ and $E[D]=\tilde{D}=D$ where $D$ is the ES-NGI diameter $=15$. In 2) once new links are added, the resulting graph remains constant in the ensemble during the tests and hence it holds again that $E[l]=l$ and $E[D]=\tilde{D}=D$. In 3), both $l(g)$ and $D(g)$ have in general a distribution whose moments depend on the graph generation mechanism used. From a battery of tests designed to gauge $l$ and $D$ variances we verified that for the random graph, according to theory, $Var(l)\rightarrow 0$, once the phase transition in connectivity at $pn_{R}=log(n_{R})$ is reached~\cite{Chung:2001hb}.
\par
Also, \textit{equivalent} graphs with the same size and approximately the same edges than ES-NGI were obtained for both the random (Erd\"os-R\'enyi model) and for the scale-free versions. These graphs are termed as ER-NGI and SF-NGI respectively. ER-NGI were built by linking $n_{R}=189$ nodes with probability $p=2ne_{R}/n_{R}(n_{R}-1)=0.02685$. The ES-NGI values of $n_{B}=5$ and $n_{C}=23$ were also used for computing services mapping. For SF-NGI networks we slightly modified the preferential attachment model~\cite{Barabasi:1999uu} by starting from $v_{0}=30$ initial vertices and by adding $e_{0}=3$ new edges per iteration through approximately $ne_{R}/e_{0}=n_{R}-v_{0}$ iterations.
\par
Once the network is obtained, non-connected solutions are filtered in order to keep the largest connected component only. Then, after setting the values of latency $LAT$ and bandwidth $BW$, the desired DCI solution is implemented as a mapping of \textit{brokers} and CEs on that network. Random, degree or betweenness based mapping algorithms are possible. Besides, proximity based strategies, where dependent jobs are preferentially mapped to closer modes, are also implemented. Application DAGs are specified through the \textit{graphML} format, where DAG structure and both node and link weights are provided. Finally, the probabilistic task allocation addressed in Sec.~\ref{sec:MODEL} is implemented resulting into an allocation vector $\bar{X}$. Once all tasks are mapped to modes, the performance metrics $MK$, $CPIC$, $SLR$ and $PE$ for every configuration point $\mathbf{x}\in\Omega$ are obtained for their statistical analysis.
	\subsection{Effect of the distributed computing infrastructure solution \label{sec:net}}
As stressed, a major factor for the performance in DCIs is the greater or lesser degree of hierarchization in their services overlay network. 
\par
\begin{figure}[h]
   		\centering
  		 \includegraphics[width=1.0\columnwidth,height=1.0\columnwidth]{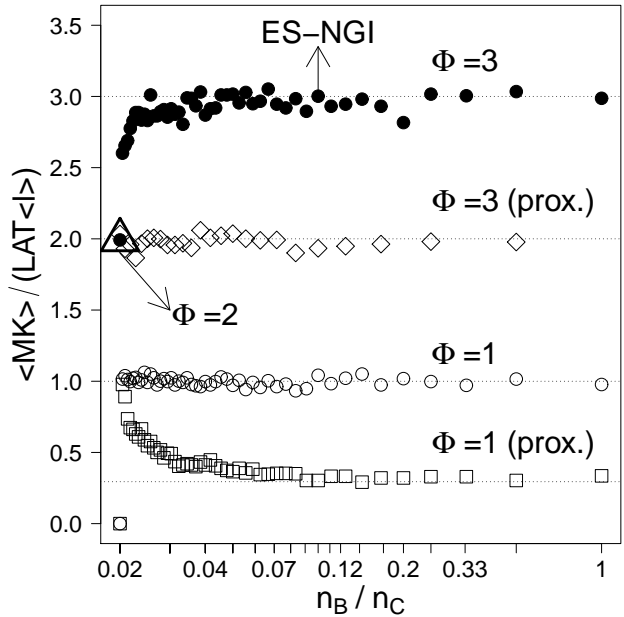}
  		 \caption{Normalised \textit{makespan} for three DCI solutions as the \textit{brokers/CE}s proportion is increased in the \textit{ER-NGI} random graph. Theoretical limits at $\Phi=1,2,3$ and $1/\langle l\rangle$ (case II in table~\ref{tab:limits}) have been also plotted as dotted lines.}
  		 \label{fig:MK_grid}
	\end{figure}

In a first experiment (Fig.~\ref{fig:MK_grid}) we monitored the average $MK/(LAT\cdot l)$ from a set of $N=1000$ samples for different network configurations and $n_{C}=50$ CEs as the amount of \textit{brokers} was ranged from $1$ to $50$ in ER-NGI. To highlight communication overheads, the allocation strategy was \textit{task repulsion} ($c=0$) and job pairs ($\omega=2$ and $c_{J}=1$) with zero processing time $PT=0$. The analytic limit for this case is $E[MK]/(LAT\cdot E[l])=\Phi$ (case II in table~\ref{tab:limits}).
In the $\Phi=2$ case only a point at $E[MK]/(LAT\cdot E[l])=\Phi=2$ coinciding with the value for the $\Phi=3$ solution for $n_{B}=1$ is obtained (in this case, the P2P solutions render $E[MK]/(LAT\cdot E[l])=0.0$, as communication overheads are zeroed). Hence, from a strictly topological perspective, it is concluded that the centralised configuration $\Phi=2$ outperforms the less hierarchical $\Phi=3$ scheme if services are mapped randomly.
\par
When a proximity optimisation is enhanced, the inter-mode \textit{broker-broker} distance is minimised and the case $\Phi=2$ is recovered. It must be stated, however, that in real DCIs, finite queuing effects at \textit{brokers} can not be neglected (in fact \textit{brokers} downfall is a common cause for DCI's malfunction). In this regard, the empirical finding of a small number of \textit{brokers} in real Grids (5 in the ES-NGI case) can be partly explained by the fact that Grids are mainly concerned in assuring job scheduling redundancy by providing $n_{B}>1$. On the other hand, P2P configurations ($\Phi=1$) significantly outperforms both the partial and total hierarchical solutions. Further, when a proximity based optimisation scheme is present, performance can be increased by augmenting the \textit{brokers} proportion. It is also noticed that this improvement stabilises in the limit $E[MK]=LAT=0.1$s that corresponds in average to a one-hop communication process (lowest dotted line at $1/E[l]$ in Fig.~\ref{fig:MK_grid}).
\par
\begin{figure}[h]
   		\centering
  		 \includegraphics[width=1.0\columnwidth,height=1.0\columnwidth]{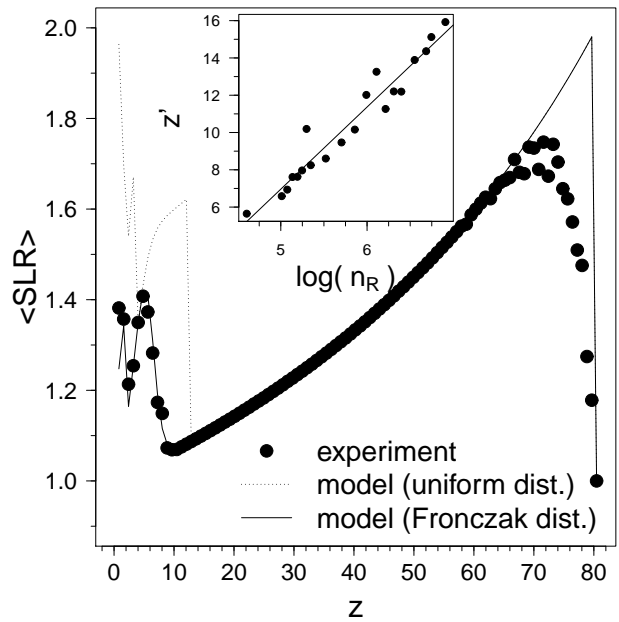}
  		 \caption{Expected value of $SLR$ using uniform and Fronczak distance distribution models for equally-sized random graphs with varying $p$. The inset shows the linear fit of the points $z'$ vs. $\log(n_{R})$ where diameter transitions from 3 to 2 have been observed.}
  		 \label{fig:phase}
	\end{figure}

Next, we focus on how communication network topology affects performance for P2P solutions. Initially, the scheduled length ratio $SLR$ is monitored under different conditions for the random graph. For every connection probability $p$ we generate $N$ graph instances of size $n_{R}$. Then we compute the corresponding averaged metrics. Whereas $l$ rapidly decreases as $p>1/n_{R}$, $D$ has, in general, a variance that depends on $p$ and $n_{R}$. In particular Bollob\'{a}s~\cite{Bollobas_RG} found that $D$ is almost surely concentrated on at most four values if $pn_{R}-\log n_{R}\rightarrow\infty$. More recently Chung and Lu~\cite{Chung:2001hb} have found how $D$ is clustered around finite sets of values for the range $1/\log(n_{R})<pn_{R}/\log(n_{R})\leq b$, where $b$ is a constant. In Fig.~\ref{fig:phase} we compare the average $SLR$ with $E[SLR]=(\tilde{D}-T_{d}(\tilde{D},c_{J}))/E[l]$ as the parameter $z=pn_{R}/log(n_{R})$ is ranged from the critical probability $p_{c}=1/n_{R}$ to $p=1$. For the distance distribution we used both the model presented by Fronczak \textit{et al.}~\cite{Fronczak:2004bh}
\begin{equation}
f_{d}(x)=1-e^{-1/n_{R}(n_{R}p)^{x}}
\label{eq:fronczak}
\end{equation}
and the uniform distribution. The model is validated through Monte Carlo tests ($N=1000$) for random graphs of size $n_{R}=500$ for $n_{J}=40$ job pairs ($\omega=2$) in a P2P solution with $n_{B}=n_{C}=50$ and $c=0$.
\par
From Fig.~\ref{fig:phase} it is noticed that the model with Fronczak \textit{et al.} distribution starts to reproduce numerical experiments for $z>2$, where the diameter is concentrated on at most 3 values~\cite{Chung:2001hb}. In this case the variance of the diameter is negligible and the approximation of~(\ref{eq:MK_GENERAL}) becomes reliable. Notice that a wide range of convergence between tests and model is found. However, this agreement is weakened as we approach the complete graph transition (the diameter abruptly changes from $2$ to $1$). On the other hand, it is well known that the distance distribution in random graphs lies far away from being uniform~\cite{Fronczak:2004bh,Blondel:2007fl} but, rather $f_{d}(x)$, shows an oscillatory behaviour with $p$~\cite{Blondel:2007fl}. However, the uniform distribution limit is interesting as it reproduces the tests for $z>12$. A detailed analysis about how the diameter transitions affect the performance is out of the scope of the present work. However, in a set of tests we monitored the diameter transitions from values 3 to 2 for different graph sizes $n_{R}$. From that study we found a linear relationship between the jumps at $z'=p'n_{R}/\log(n_{R})$ and $log(n_{R})$ (see the linear fit in Fig.~\ref{fig:phase} inset). For $n_{R}=500$ the linear fit renders a value of $z'=12.3$ that corresponds with the point $z$ in Fig.~\ref{fig:phase} where the uniform distribution model starts to reproduce experiments.
\par
It should be noted the local minimum found at $z=10$. In this case a random mapping of tasks (neglecting topological and smart service mapping strategies) would render an optimal performance. If the ES-NGI topology was random, this would involve a connection probability of $p=0.277$ (a factor of $10.66$ times higher than the  connection probability of \textit{ER-NGI}). From this value, $D$ keeps constant while $l$ smoothly decreases  for a wide range or $z$ and, hence,  $SLR$ increases. This trend is inverted near the complete graph limit $(z>65)$ where $D$ falls sharply to $1$.
In this region the model is unable to reproduce the last unstable values of $D$ (peak point at  $SLR\approx 2$).	
\par
\begin{figure}[h]
   		\centering
  		 \includegraphics[width=1.0\columnwidth,height=1.0\columnwidth]{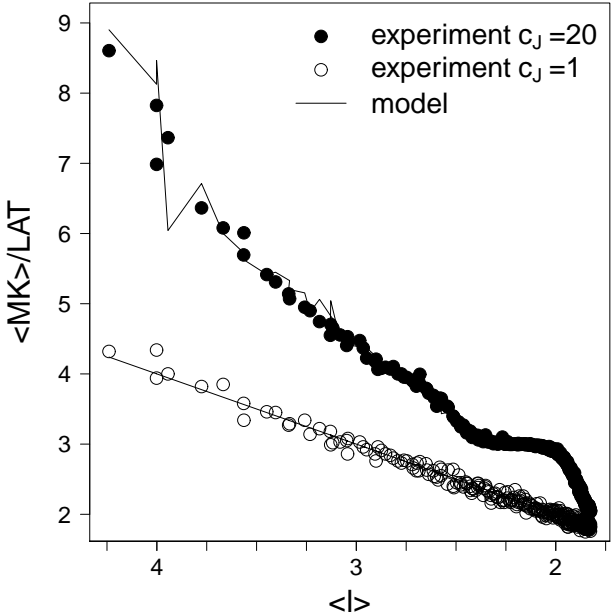}
  		 \caption{Numerical validation for the model for $E[MK]$ as a function of the average path length in the empirical ES-NGI topology for $c_{J}=20$  and $c_{J}=1$. In the latter case the linear limit $E[MK]/LAT=l$ (case II in table~\ref{tab:limits}) is reproduced.}
  		 \label{fig:MK_r}
	\end{figure}

In another set of tests we gauge the empirical ES-NGI \textit{makespan} as new links are randomly added. We allocated job pairs with $PT=0$ on a $\Phi=1$ configuration ($n_{B}=n_{C}=50$ randomly mapped modes) with $c=0$. These experiments were also designed to validate the model in the case of an empirical distance distribution~(\ref{eq:MK_c0}). In Fig.~\ref{fig:MK_r} we compare $\langle MK\rangle/LAT$ obtained from numerical simulations ($N=500$ samples) with $E[MK]/LAT=E[\tilde{D}-\sum_{x=0}^{\tilde{D}-1}[f_{d}(x)]^{c_{J}}]$ for $c_{J}=20$ and $c_{J}=1$. In this case $f_{d}(x)$ was estimated through the empirical cumulative distribution function from the ES-NGI graph.
For $c_{J}=1$ it can be noticed that the linear relationship $E[MK]/LAT=l$ (case II in table~\ref{tab:limits}) is reproduced (solid line in Fig.~\ref{fig:MK_r}).
\par
\begin{figure}[h]
   		\centering
  		 \includegraphics[width=1\columnwidth,height=1\columnwidth]{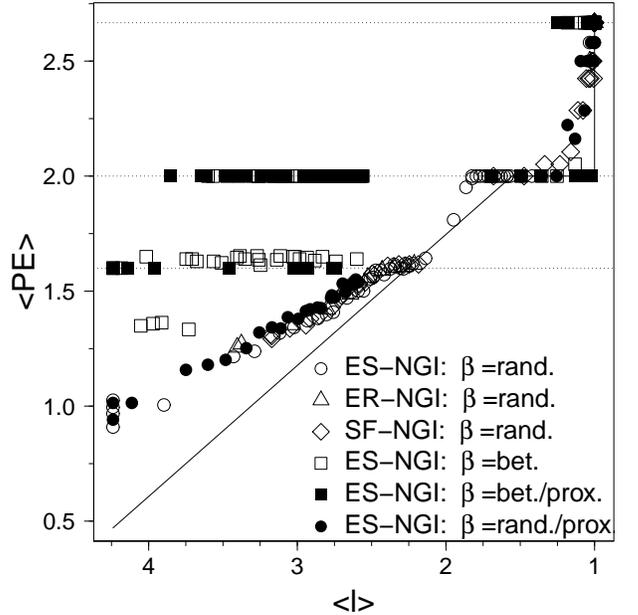}
  		 \caption{Average parallel efficiency for ES-NGI and its random and scale-free equivalents as the average path length decreases. Results from both random and betweenness based computing services mappings and task allocations through a proximity criterium are shown. The solid line corresponds to the value of $E[PE]$ in case III of table~\ref{tab:limits} for $\Phi=1$ and  the dotted horizontal lines correspond to that value when $\tilde{D}=1,2,3$.}
  		 \label{fig:PEequivs}
	\end{figure}

Finally, we explored how the parallel efficiency of the ES-NGI could be improved by allowing more efficient computing services mappings. For a P2P solution ($\Phi=1$) with $c=0$ the expression for $E[PE]$ (neglecting second order terms in the variance of $MK$) renders
\begin{equation}
E[PE]=\frac{c_{J}/n_{S}}{1+\frac{2\eta}{\omega} [\tilde{D}(\omega-1)-\sum_{x=0}^{\tilde{D}(\omega-1)-1}[f_{D^{bb}(x)}]^{c_{J}}]}.
\label{eq:pe}
\end{equation}
We launched $c_{J}=20$ job pairs ($\omega=2$) in a \textit{RoundRobin} ($c=0$) fashion with $\eta=1/2$. The expression for $E[PE]$ from table~\ref{tab:limits}  is a lower bound of~(\ref{eq:pe})) (solid line in Fig.~\ref{fig:PEequivs}). In this regard, the \textit{efficiency} of a topology in a P2P network is driven by the quantity: $\sum_{x=0}^{\tilde{D}-1}[f_{D^{bb}(x)}]^{c_{J}}$. The higher this value, the more efficient P2P infrastructures are obtained. In Fig.~\ref{fig:PEequivs} we show $\langle PE\rangle$ in the ES-NGI and its random, and scale-free equivalents when the average shortest path $l$ is decreased by adding new links. For these tests Monte Carlo simulations from samples with $N=500$ size were conducted for different computing services mappings and task allocation algorithms. 
As it was found in~\cite{daFontouraCosta:2005wn}, the parallel efficiency increases by adding new links randomly (i.e., decreasing $l$) for both random and scale-free graphs. This is also confirmed in our tests for both the empirical ES-NGI Grid and for its synthetic counterparts.
\par
We also monitored the efficiency for \textit{smarter} mappings of computing services. In particular it was found that when \textit{brokers} were preferentially mapped to nodes with higher betweenness connectivity ($\beta=bet.$ in Fig.~\ref{fig:PEequivs}), $\langle PE\rangle$ increased significantly with respect to the random mapping version. Also, by providing a distance optimisation mechanism (tasks are first allocated to closer modes) these results could be further improved. The abscissas (dotted lines in Fig.~\ref{fig:PEequivs}) correspond to the limit case of $E[PE]$ in table~\ref{tab:limits} for uniform distributions $E[PE]=(c_{J}/n_{S})/(1+\eta\tilde{D})$ at values $\tilde{D}=1,2,3$. It can be noticed how the betweenness based mappings cluster around these limits. In this situation the contribution of the graph distance distribution to the efficiency is not significant. Finally, the maximum efficiency is clearly achieved for the complete graph when $\tilde{D}=1$.
	\subsection{Effect of the application workflow structure \label{sec:dag} and allocation strategy}	
\begin{figure}[h]
   		\centering
  		 \includegraphics[width=1\columnwidth,height=1\columnwidth]{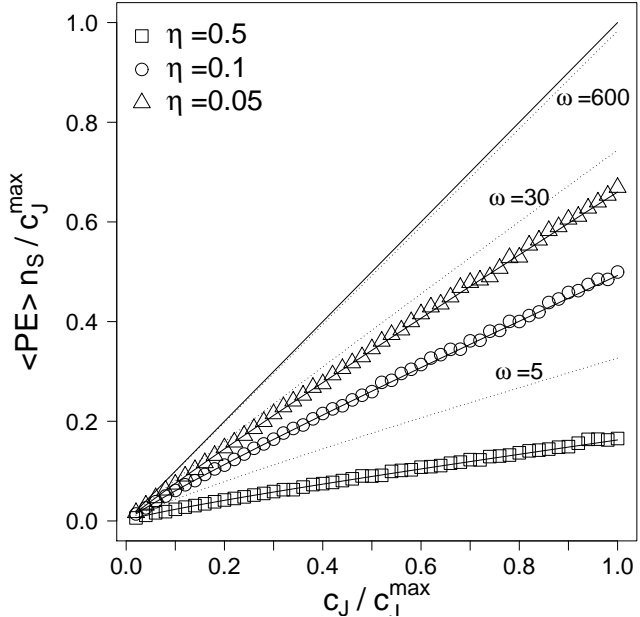}
  		 \caption{Normalised parallel efficiency vs. task cluster number. Solid lines represent the model from~(\ref{eq:pe}) for $\omega=2$. Symbols represent the results from numerical tests and dotted lines represent numerical solutions for $\omega=5,30,600$ with $\eta=0.5$. The upper limit in efficiency (diagonal solid line) corresponds to $E[PE]n_{S}/c_{J}^{max}=c_{J}/c_{J}^{max}$.}
  		 \label{fig:cj}
	\end{figure}

We shall now examine how the application workflow parameters ($c_{J}$, $\omega$) and the job clustering $c$ affect the performance metrics. Firstly, we monitor (Fig.~\ref{fig:cj}) the normalised parallel efficiency $E[PE]n_{S}/c_{J}^{max}$ in the ES-NGI topology as the normalised number of job clusters $c_{J}/c^{max}_{J}$ increases. A P2P scheme with $n_{S}=150$ randomly mapped modes and $c=0$ was used.
As expected from~(\ref{eq:pe}) (an equivalent result was reported in~\cite{daFontouraCosta:2005wn}) the parallel efficiency grows with the number of task clusters.  An upper bound is found of $E[PE]n_{S}/c_{J}^{max}=c_{J}/c_{J}^{max}$ (diagonal solid line in Fig.~\ref{fig:cj}). According to~(\ref{eq:pe}) this limit is reached when $\eta\rightarrow 0$ or when $\omega>>1$ for a finite $\eta$ value. Conversely, if $c_{J}$ is small there will be not enough computational work to distribute in the computing modes resulting in small efficiencies. Also in da Fontoura \textit{et al.}~\cite{daFontouraCosta:2005wn} it is claimed that efficiency increases with the task execution time. In our case this has been also observed when $\eta$ decreases; larger values for $PE$ are obtained for big processing times and low latency networks (see Fig.~\ref{fig:cj}). 
\par
\begin{figure}[h]
   		\centering
  		 \includegraphics[width=1\columnwidth,height=1\columnwidth]{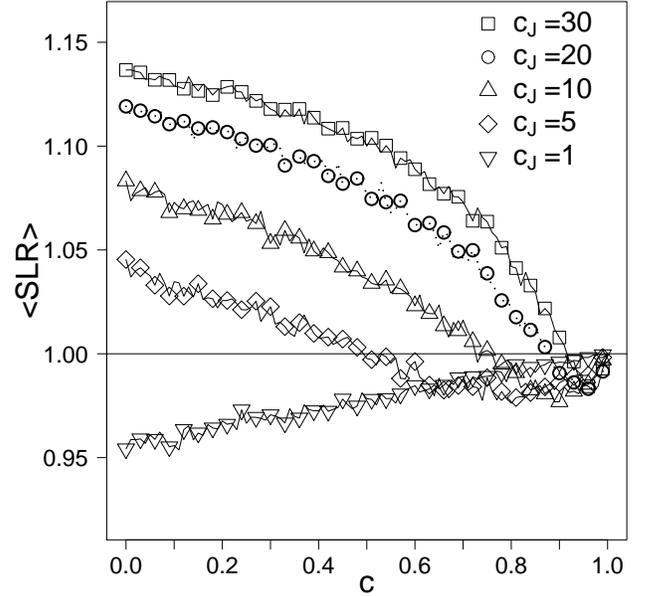}
  		 \caption{Average $SLR$ for different cluster numbers as the task clustering parameter increases. The monitored DCI is the ES-NGI topology in a P2P solution when services are mapped through a betweenness based optimisation scheme.}
  		 \label{fig:cluster}
	\end{figure}

In the last set of experiments we investigate how $SLR$ depends on the task clustering parameter $c$. For this case from~(\ref{eq:E_CPIC_w2}) and~(\ref{eq:MK_OPERATIVA}) it is found
\begin{equation}
E[SLR]=\frac{1+\eta[\tilde{D}-\sum_{n=0}^{c_{J}}P_{1-c}(c_{J},n)\sum_{x=0}^{\tilde{D}-1}(f_{d}(x))^{n}]}{1+\eta E[l](1-c^{c_{J}})}
\label{eq:E_SLR_c}
\end{equation}
with the limiting case of $SLR=1$ for $c_{J}=1$. In Fig.~\ref{fig:cluster} we monitor the numerical average of $SLR$ ($N=200$ samples) for the ES-NGI graph in a P2P solution of $n_{S}=50$ modes deployed through a betweenness optimisation scheme by allocating job pairs with different clusterings and $\eta=0.5$.
On the left side ($c=0$) dependent jobs are allocated in a \textit{RoundRobin} fashion and communication overheads are maximum. As $c$ increases tasks tend to be allocated in the same computing modes and those communication costs decrease. In the right side  for $c=1$, all tasks are mapped to the same computing mode which, in average, has a distance of $D^{cc}\approx l$ and, hence $E[SLR]=1$. Interestingly, lower values for $E[SLR]$ than the minimum achievable in a random services mapping (solid line corresponding to $E[SLR]=1$ in Fig.~\ref{fig:cluster}) can be obtained for different $c$ and $c_{J}$ combinations. 
Furthermore, from $c>0.6$, the strategy for $c_{J}=5$ outperforms the solution for $c_{J}=1$; the allocation strategy \textit{scales} with the number of tasks and the higher $c_{J}$ the better task allocation is achieved with respect to the \textit{random} mapping.
This evidences that when more accurate services mappings in a distributed computing infrastructure are addressed, optimised strategies can be enhanced by coupling task allocation and topology.  However, such analysis is out of the scope of the present contribution and is left for future research. 
\section{Conclusion and final remarks \label{concl}}
This work has been aimed at finding analytical expressions for the expected values of distributed computing performance metrics when the communication network has a complex structure. Through active \textbf{traceroute}-like probing tests a \textit{seed} network was built from a real computing Grid. From this graph, ensembles of synthetic graphs with additional structure were generated to verify the model and to monitor different scenarios through numerical simulation.  Although inspired by the Grid case, the models and results reported can be easily extended without loss of generality to other DCI solutions. Finally, to the authors' knowledge, this is the first work that provides analytical expressions for the expected performance metrics when job dependencies and a complex network structure are considered. The main results obtained are summarised below.
\begin{enumerate}
\item Centralised schemes in distributed computing infrastructures in general increase parallel applications execution time due to the extra communication overheads between computing services. The optimal solution corresponds to a P2P model ($\Phi=1$).  On the other hand, if no additional optimisation strategies in the services mappings or task allocations are enhanced, common Grid infrastructures ($\Phi=3$) do not improve performance when the \textit{brokers/CE}s proportion increases. Moreover, the totally centralised ($\Phi=2$) solution outperforms the ES-NGI Grid, although this improvement can also be achieved when tasks are preferentially mapped to closer resources.
\item By addressing more \textit{intelligent} computing services mappings (in particular, a preferential mapping based on betweenness centrality  or if tasks are more likely to reach closer modes), significant improvements in the parallel efficiency and scheduled length ratio are achieved. For random and scale-free equivalent graphs these improvements are less appreciable.
\item When the \textit{seed} topologies are incrementally modified by the addition of new edges, the analytical expressions for the expected values of $MK$ and $SLR$ are numerically reproduced in a wide range of graphs. For the random graph, a local optimum is found in $SLR$ meaning that if distributed computing topologies hold a random-like communication network structure, targeted reconnections could improve the infrastructure performance significantly. 
\item Once the ES-NGI topology has been enriched with a mode betweenness based mapping, an optimisation in the parallel efficiency can be achievable by coupling distance distribution $f_{d}(x)$ and task clustering parameter $c$. In this case, solutions with higher number of task clusters outperforms the performance of a single job pair ($c_{J}=1$).  	
\item The parallel efficiency increases for denser graphs and for larger task clusters and processing times. This confirms qualitatively the results from~\cite{daFontouraCosta:2005wn}.
\end{enumerate}
Finally, we make some remarks on the scope and assumptions of the addressed model. Firstly, a \textit{Virtual} node based communications network (where every node does not necessarily correspond to a communication device) with symmetrical paths and minimal distance based routing was considered. Further, the communication network was regarded as homogeneous in both latency and bandwidth. Background traffic effects were also neglected.
\par
The computing elements were considered as uniform in processing speed and with no additional queuing delays. It was also assumed that \textit{brokers} did not introduce any delay when mapping tasks. This constitutes a threat for the reliability of the model since, on the one hand, the local queue effect is a major factor in the performance of DCIs (and Grids in particular)~\cite{Muttoni:2004us} and, on the other hand, a notable characteristic of global DCIs is their heterogeneity in resources~\cite{Foster:2001th}. In a future we will investigate the coupling of those effects in the addressed model.
\par
Moreover, no additional ordering is imposed to tasks when they reach a CE: they are processed in parallel if no precedence relationship exists between them. This slightly deviates our model from the traditional scheduling approach where both allocation and order should be specified.
 \par
 We considered a \textit{minimalist} DAG model with equally sized jobs and transfer files arranged into task clusters with the same dimension. Further, job sizes were regarded as small compared with the network bandwidth (i.e., \textit{latency-bounded} approximation).
\par	
While these assumptions may turn the addressed model less reliable, it is our belief that including analytical expressions that link topology, application structure and allocation strategy parameters may render the distributing computing problem more manageable, easing the incorporation of additional effects. Furthermore, the techniques addressed in this work (e.g., the application of order statistics analysis to DCI scheduling) have been validated numerically and are amenable to be used with real data.
\acknowledgments
CETA-CIEMAT acknowledges the support received from the European Regional Development Fund through its Operational Program \textit{Knowledge-based Economy}. A. A. acknowledges support from the Ministerio de Ciencia e Innovaci\'{o}n (Spain) through Grant No. FIS2010-16587 and from the Junta de Extremadura (Spain) through Grant No. GR10158, partially financed by FEDER (Fondo Europeo de Desarrollo Regional) funds.
\appendix
\numberwithin{equation}{section}
\section{Probabilistic job allocation model\label{ap0}}
From the conditions addressed in Sec.\ref{sec:MODEL} the following simple model is proposed
\begin{equation}
p_{kn|lm}=A_{kl}\Gamma_{kl}+(1-A_{kl})p_{lm}.
\label{eq:AP_MODEL_COND}
\end{equation}
Here $\Gamma_{kl}$ is a function of $n,m$ to be determined. By imposing normalisation conditions it holds: 1) $\Gamma_{kl}\in[0,1]$ and 2) $\sum_{m}\Gamma_{kl}=1$. We also require that $\Gamma_{kl}=c\in[0,1]$ if $m=n$  being $c\in[0,1]$ defined as the  \textit{job clustering parameter}. By using the former conditions over $\Gamma_{kl}$ 
\begin{equation}
p_{kn,lm}=\frac{A_{kl}\cdot p_{kn}}{n_{S}-1}[\delta_{nm}(c\cdot n_{S}-1)+1-c]+(1-A_{kl})p_{kn}\cdot p_{lm}.
\label{eq:AP_MODEL}
\end{equation}
We also define the \textit{clustering probability} $\tilde{p}_{kl}=P[\Delta_{kl}=1]=\sum_{n}p_{kn,ln}$ as the probability of jobs $J_{k}$ and $J_{l}$ are mapped into the same mode. By assuming also $S$ as equiprobable ($p_{kn}=1/n_{S}\;\forall k$)~(\ref{eq:AP_MODEL}) renders
\begin{equation}
\tilde{p}_{kl}=c\cdot A_{kl}+\frac{(1-A_{kl})}{n_{S}}.
\label{eq:P_delta}
\end{equation}
\section{Derivation of analytical expressions for limit cases\label{ap}}
\begin{enumerate}
\item job-pairs limit. In this case, a single clustering variable exists $W=1-\Delta$ and $f_{W}(0)=P[1-\Delta\leq 0]=P[\Delta=1]=c$.
Then,~(\ref{eq:CPIC_GENERAL}) reduces to
\begin{equation}
E[CPIC]=2PT+\Phi\cdot LAT\cdot E[l](1-c^{c_{J}}).
\label{eq:E_CPIC_w2}
\end{equation} 
Further $U=(1-\Delta)D^{cc}$, and, hence
\begin{eqnarray}
\nonumber E[MK]=2 PT+\\
LAT[\Phi\tilde{D}-\sum_{x=0}^{\Phi\tilde{D}-1}(f_{(1-\Delta)D^{cc}}(x))^{c_{J}}].
\label{eq:E_MK_w2}
\end{eqnarray}
It is noticed that for the computation of $f_{(1-\Delta)D^{cc}}(x)$ a product form for $S^{n_{J}}=S^{\omega\cdot c_{J}}=(S^{2})^{c_{J}}$ can be used through the partition: $S^{2}=\{(X_{i},X_{i+1})\}$, where the sample space is split into overlapping $"\Delta=1"=\{(X_{i},X_{i+1})\in S^{2}: X_{i}=X_{i-1}\}$ and non-overlapping events. 
Since it also holds that $"\Delta=1"\subset "D^{cc}\leq x",\forall x$, the event $"(1-\Delta)D^{cc}\leq x"$ can be rewritten as $"\Delta=1"\cup("\Delta=0"\cap"D^{cc}\leq x")$. As a consequence, distribution functions can be expressed as
\begin{equation}
f_{(1-\Delta)D^{cc}}(x)=f_{D^{cc}}(x)+c\tilde{f}_{D^{cc}}(x).
\label{eq:f_U}
\end{equation}
\item Random mapping of computing services. In this case $D^{xy}, xy\in\{cb,bb,bc\}$ are IID simple random variables with distribution $f_{d}(x)$. Then $f_{D^{xy}}(x), xy\in\{cb,bb,bc\}$ can be thought as the empirical distribution functions from samples of size $n_{B}$ and $n_{C}$. If $n_{B}$ and $n_{C}$ are large enough compared with $n_{R}$, by Glivenko-Cantinelli theorem~\cite{kendall1945advanced} it holds that $f_{D^{xy}}(x)$ converges to $f_{d}(x), xy\in\{cb,bb,bc\}$. Furthermore, as in this case \textit{brokers} and CEs can be interchanged, by symmetry we set that: $D^{bb}\approx D^{cb}\approx D^{bc}$, and, hence, $D^{cc}\approx \Phi D^{xy},xy\in{bb,bc,cb}$. As a consequence, $\forall xy\in{bb,cb,bc}$
\begin{equation}
f_{D^{cc}}(x)\approx f_{D^{xy}}(x/\Phi)\rightarrow f_{d}(x/\Phi)=f_{\Phi d}(x).
\label{eq:Dcc}
\end{equation}
In other words, the random variable $U$ can be thought as a rescaling of $d$ in a factor of $\Phi$. In this case~(\ref{eq:E_MK_w2}) by using~(\ref{eq:f_U}) and~(\ref{eq:Dcc}) renders
 \begin{eqnarray}
 E[MK]=2 PT +LAT(\Phi\tilde{D}-\\
 \nonumber \sum_{x=0}^{\Phi\tilde{D}-1}[f_{\Phi d}(x)+c\tilde{f}_{\Phi d}(x)]^{c_{J}}).
 \label{eq:E_MK_UNIF}
 \end{eqnarray}
 \end{enumerate}
 As $f_{\Phi d}(x)+\tilde{f}_{\Phi d}(x)=1$, it is noticed that
 \begin{eqnarray}
 \nonumber E[MK]=2PT+\\
 LAT[\Phi\tilde{D}-\sum_{n=0}^{c_{J}}P_{1-c}(c_{J},n)T_{\Phi d}(\Phi\tilde{D},n)],
 \label{eq:MK_OPERATIVA}
 \end{eqnarray}
where through the binomial theorem $P_{1-c}(c_{J},n)=\binom{c_{J}}{n}(1-c)^{n}c^{c_{J}-n}$ is the success probability of obtaining $n$ successes in a Bernoulli trial with probability $1-c$ and $T_{\Phi d}(\Phi\tilde{D},n)=\sum_{x=0}^{\Phi\tilde{D}-1}(f_{\Phi d}(x))^{n}$. It is noticed that $P_{0}(c_{J},n)=\delta_{n,0}$ and $P_{1}(c_{J},n)=\delta_{n,c_{J}}$. Further, from~(\ref{eq:l_tail}) and~(\ref{eq:Dcc}) it can be found that  $T_{\Phi d}(\Phi\tilde{D},0)=\Phi\tilde{D}$ and $T_{\Phi d}(\Phi\tilde{D},1)=\Phi(\tilde{D}-E[l])$.
\newpage
Two limit cases are investigated:
 \begin{itemize}
 \item $c=1$ (\textit{total task clustering}), where $E[CPIC]=E[MK]=2PT$
 \item $c=0$ (\textit{total task repulsion}). In this case $E[CPIC]=2PT+LAT\cdot \Phi\cdot E[l]$ and 
 \begin{equation}
 E[MK]=2PT+LAT[\Phi\tilde{D}-T_{\Phi d}(\Phi\tilde{D},c_{J})].
 \label{eq:MK_c0}
 \end{equation}
 Two sub-cases are considered:
 \begin{enumerate}
 \item $c_{J}=1$.
 Now a linear relationship between $E[MK]$ and $E[l]$ is found and
 \begin{equation}
 E[MK] = 2 PT +\Phi\cdot LAT\cdot E[l].
 \label{eq:MK_cj1}  
 \end{equation}
\item $c_{J}>>1$ and uniform distance distribution. 
If $f_{d}(x)$ is also regarded as uniform, from~(\ref{eq:Dcc}) $f_{\Phi d}(x)=x/(\Phi\tilde{D})$ and it is obtained
\begin{equation}
T_{\Phi d}(\Phi\tilde{D},c_{J})=\frac{1}{(\Phi\tilde{D})^{c_{J}}}\sum_{x=0}^{\Phi\tilde{D}-1}x^{c_{J}}.
\label{eq:F_UNIF}
\end{equation}
In the limit $c_{J}>>1$ this term is neglected in~(\ref{eq:MK_c0}) and the expression for  $E[MK]$ renders
\begin{equation}
E[MK]=2PT+LAT\cdot \Phi\tilde{D}.
\label{eq:MK_cj1}
\end{equation}
 \end{enumerate}
 \end{itemize}
 \par
\ed